# Preventing overfitting in infrared ellipsometry using temperature dependence: fused silica as a case study

**Shenwei Yin[1,†], Jin-Woo Cho[1,†], Demeng Feng[1], Hongyan Mei[1], Tanuj Kumar[1], Chenghao Wan[1], Yeonghoon Jin[1], Minjeong Kim[1], and Mikhail A. Kats[1*]**

[1]*Department of Electrical and Computer Engineering, University of Wisconsin-Madison, Madison, Wisconsin 53706, USA.*
[†]*These authors contributed equally.*
**mkats@wisc.edu*

## Abstract

Fitting oscillator models to variable-angle spectroscopic ellipsometry (VASE) data can lead to non-unique, unphysical results. We demonstrate using temperature-dependent trends to prevent overfitting and ensure model physicality. As a case study, we performed mid-infrared VASE measurements on fused silica ($SiO_2$) of various grades, from room temperature to 600 °C. We fitted oscillator models independently at each temperature, and confirmed the model's physical validity by observing the expected monotonic trends in vibrational oscillator parameters. Using this technique, we generated a highly accurate dataset for the temperature-dependent complex refractive index of fused silica for modeling mid-infrared optical components such as thermal emitters.

## 1. Introduction

Spectroscopic ellipsometry is a powerful technique to measure the complex refractive index of thin films and bulk materials [1–3], with ellipsometry instruments available in the ultraviolet [4], visible [5], infrared [5,6], and terahertz ranges [7]. In ellipsometry, light is incident at one or several oblique angles onto a sample with a certain polarization, and the polarization state of the specular reflected light is measured. The spectroscopic measurement can be performed either one wavelength at a time using a monochromator [8], or via a Fourier-transform approach where wavelengths are multiplexed [9]. The former method is typically used for ultraviolet, visible, and near-infrared wavelengths, while the latter is more common in the mid- to far-infrared range. These measurements are then used to fit unknown material parameters such as the complex refractive index of one or more layers, and/or the thickness of the layers. In particular, when the goal is to determine the complex refractive index of some material, the standard practice is to build a model for the permittivity (and hence the refractive index) that is Kramers-Kronig consistent [10,11], and therefore returns a physical solution.

The goal of a high-quality spectroscopic ellipsometry experiment is to obtain enough measurements such that the resulting fit is unique—i.e., only one combination of refractive index (or multiple indices and thicknesses in the case of multilayer materials) can fit the experimental data. Uniqueness can be achieved by measuring over a sufficiently broad wavelength range, multiple angles of incidence, and/or performing auxiliary measurements such as optical reflectance and transmittance or non-optical measurements such as creating a step in the film (by etching or liftoff) and measuring film thickness with a contact profilometer or an atomic force microscope. Nevertheless, finding a unique model can be a challenge. Furthermore, even if a fitted model is unique in terms of the values of the refractive index and thickness, it may not provide sufficient physical intuition about, e.g., the vibrational and electronic states in the material, because many different models may result in very similar spectra of the complex refractive index [12,13].

In this work, we demonstrate that temperature-dependent variable-angle spectroscopic ellipsometry (VASE) can enable the identification of a unique and physically meaningful model for a material. Our test case is wafers of various grades of fused silica (amorphous $SiO_2$), measured in the mid infrared, for wavelengths of 5 to 25 μm, where we anticipate that the resulting precisely measured temperature-dependent optical properties will be useful for designing mid-infrared thermal emitters [14,15] or radiative coolers [16,17] and other optical components [18–20]. In the mid-infrared range, fused silica has multiple vibrational resonances which should be described as oscillators in models of its optical properties [15,21–25]. As temperature increases, thermal expansion (which impacts bond stiffness) [26,27] and anharmonicity (which arises from phonon-phonon interactions) [28] leads to predictable, monotonic shifts in these vibrational resonances, affecting the complex refractive index. By performing VASE measurements over a range of temperatures, we can utilize this expected monotonic temperature dependence to confirm whether the model we selected is physical, and to prevent overfitting. Using temperature-dependent VASE, we generated a highly accurate dataset of the temperature-dependent mid-infrared properties of various grades of fused silica.

## 2. Ellipsometry measurements

The basic variable-angle spectroscopic ellipsometry (VASE) measurement and our test materials are described in Fig. 1a and Table 1. Our samples are wafers of fused silica from Corning with various levels of metallic impurities and OH content. "7980 Standard grade" fused silica has < 100 parts per billion (ppb) metallic impurities and 800–1000 parts per million (ppm) OH content [29], which is sufficient for most optical applications requiring high transparency in the visible [30] and near infrared [31]. Fused silica can also be made more transparent in the infrared (IR) or ultraviolet (UV) wavelengths by further purification to minimize impurity absorption; two higher grades are shown in Table 1. Though we do not expect metallic impurities or OH content at the ppm or ppb level to significantly affect the mid-IR optical properties [18,20,29,31], measurement of these minor differences can serve as a good benchmark for our technique and are explored later in this paper.

All our samples are two-inch wafers with thickness of 0.5 mm. We sandblasted the backside of the samples to minimize back reflections [1–3,5], and limited our reporting wavelength to > 5 μm, where fused silica is absorbing due to the presence of several vibrational resonances [21,22].

**Table 1. List of samples characterized in this work, with impurity data from the Corning spec sheet [29].**

| Sample name | Metallic impurities (ppb) | OH content (ppm) |
|---|---|---|
| 7980 Standard Grade | < 1000 | 800–1000 |
| 7979 IR Grade | < 100 | < 1 |
| 8655 ArF Grade | < 10 | < 1 |

Our VASE measurements were performed using a Fourier-transform-type ellipsometer (J. A. Woollam IR-VASE) at incidence angles of 50, 60, and 70° [Fig. 1a]. This range of incidence angles is commonly used in VASE to span the angular range from below to above the Brewster angle, to maximize the signal-to-noise ratio of the raw experimental data. For absorbing materials like fused silica in the mid-IR, one can calculate a wavelength-dependent pseudo-Brewster angle, which is the angle at which the p-polarized reflectance is minimized [32–34]. Our 50–70° VASE measurement spans most of the pseudo-Brewster angles in the measurement wavelength range [Fig. S1].

The depolarization values that constitute raw data from ellipsometry measurements are usually reported as wavelength-dependent values Ψ and Δ, which are related to the complex-valued reflectance coefficients [1,3]:

$$\frac{r_p}{r_s} = \tan(\Psi)\, e^{i\Delta}, \tag{1}$$

where $r_p$ is the complex-valued reflectance of the sample for $p$-polarized light, and $r_s$ is the complex-valued reflectance for $s$-polarized light. The measured Ψ and Δ for standard-grade fused silica at room temperature are shown in Figs. 1b and c.

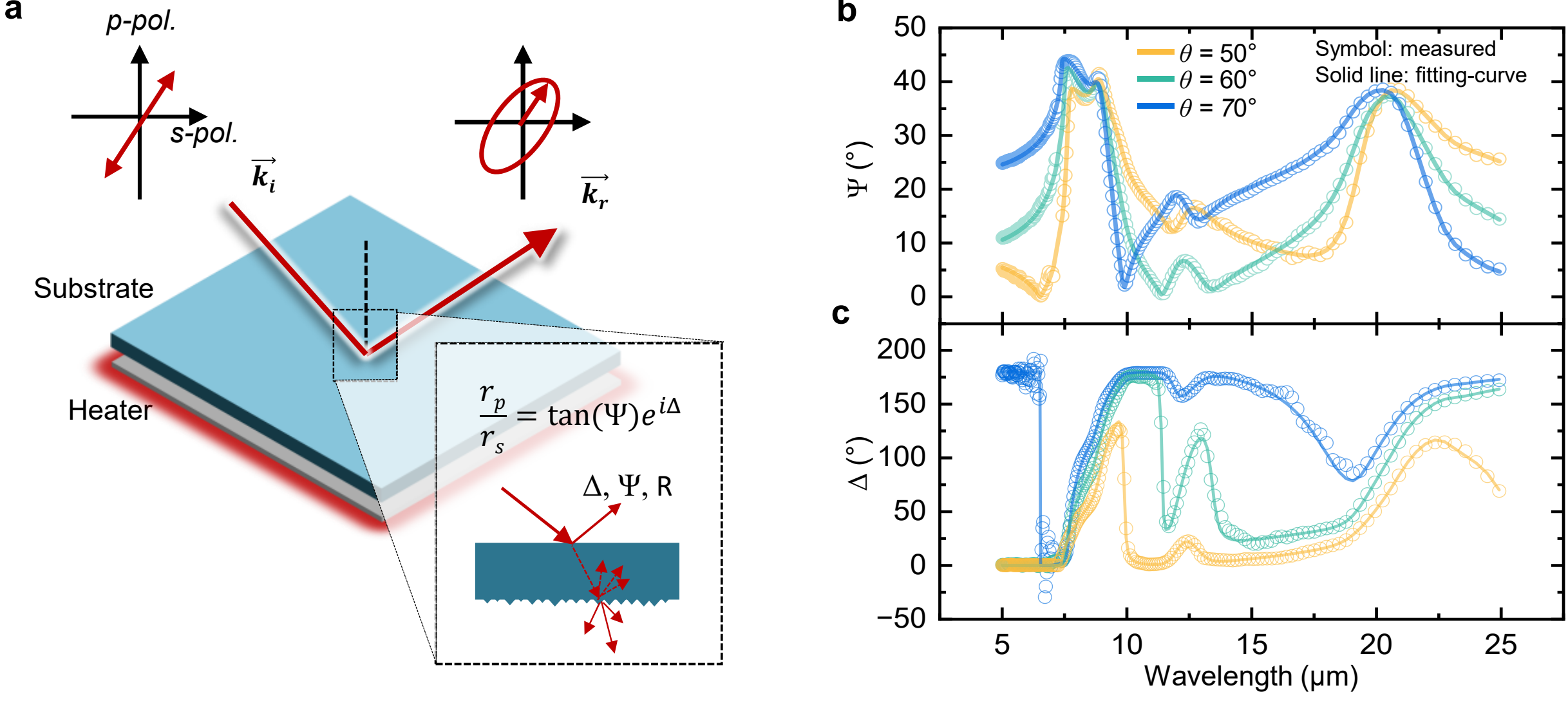

Figure 1. (a) Schematic of our temperature-dependent spectroscopic ellipsometry experiment, showing incident/reflected beams and the definitions of Ψ and Δ. (b,c) Experimental Ψ and Δ data (symbols) and corresponding fitted result (lines) using 6 Gaussian oscillator at incidence angles of 50°, 60°, and 70° for standard-grade fused silica (Corning 7980) at room temperature (i.e., 24 °C). The details of the fitting will be discussed later.

Since one of the goals of this work is to generate highly accurate and precise datasets, for each grade of $SiO_2$, we performed temperature-dependent VASE using a temperature-controlled stage (Linkam Scientific, Model HFSEL600). We performed a full ellipsometry measurement at several temperatures: 24, 100, 200, 300, 400, 500, and 600 °C. Prior to each measurement, we waited 15 minutes after reaching the target temperature. The ellipsometric parameters (Ψ and Δ) of each sample at each temperature were measured for wave numbers between 400 and 2000 $cm^{-1}$ (corresponding to wavelengths between 5 and 25 μm) with a resolution of 16 $cm^{-1}$ [specific ellipsometer settings described in Supplement Section I]. Fitting at each temperature for each sample was performed independently and manually using the WVASE software.

## 3. Results and Discussions

The ellipsometry measurements of fused-silica wafers (i.e., spectra of $\Psi$ and $\Delta$ in Figs. 1b and c) can in principle be fitted using different Kramers-Kronig-consistent models, such as combinations of Lorentzian, Gaussian, and other oscillator types [1,3]. We attempted to fit the data using a combination of Lorentzian oscillators, and separately using a combination of Gaussian oscillators, and also mixing the two oscillator types, and found that we could obtain better fits using only Gaussian oscillators [Fig. S3]. This is consistent with the use of Gaussian oscillators in the literature [25,35,36], and is likely due to inhomogeneous broadening of vibrational modes in the amorphous structure of the material [37–39].

We used the following form of Gaussian oscillators, here as a function of spectroscopic wavenumber [40]. For one of the Gaussian oscillators in the model, labeled m:

$$\varepsilon_{Gaussian}(E) = \varepsilon_1(E) + i\varepsilon_2(E) \tag{2}$$

$$\varepsilon_2(E) = A_m e^{-\left(\frac{E-E_m}{\sigma_m}\right)^2} - A_m e^{-\left(\frac{E+E_m}{\sigma_m}\right)^2}, \tag{3}$$

$$\varepsilon_1(E) = \frac{2}{\pi} P \int_0^{\infty} \xi \frac{\varepsilon_2(\xi)}{\xi^2 - E^2} d\xi, \tag{4}$$

$$\sigma = \frac{Br_m}{2\sqrt{\ln(2)}}, \tag{5}$$

where $\varepsilon_1$ is the real part of the dielectric function, $\varepsilon_2$ is the imaginary part of the dielectric function, $A_m$ is the amplitude of the oscillator (dimensionless), $Br_m$ is the broadening of the oscillator ($cm^{-1}$), $E_m$ is the spectral position of the oscillator in spectroscopic wavenumber ($cm^{-1}$), and $E$ is the wavenumber of light ($cm^{-1}$). $P$ is the Cauchy principal value of the integral, $\xi$ is the integration variable in wavenumber. Eqn. (3) defines a Gaussian at both positive and negative frequencies to maintain the odd symmetry for $\varepsilon_2$, as required by Kramers-Kronig criteria [41]. Eqn. (4) ensures the real and imaginary parts of the dielectric function are Kramers-Kronig consistent [42].

We can obtain the complex refractive index by summing the contributions of all the Gaussian oscillators and $\varepsilon_\infty$, which is the value of the dielectric function at frequencies much higher than the highest-frequency oscillator:

$$\tilde{n}^2 = (n + i\kappa)^2 = \varepsilon_\infty + \sum_m \varepsilon_{Gaussian}, \tag{6}$$

where $\tilde{n}$ is the complex refractive index with real part $n$ and imaginary part $\kappa$. Thus, our fitting parameters for each material are $A_m$, $Br_m$, and $E_m$ for each oscillator (labeled $m$), and $\varepsilon_\infty$.

In the literature, there are descriptions of four vibrational modes that are prominent in fused silica: asymmetric stretching of Si-O-Si at around 1070–1120 $cm^{-1}$, stretching of Si-OH at around 950 $cm^{-1}$, symmetric stretching or bending of Si-O-Si at around 800 $cm^{-1}$, and bending (rocking) of Si-O-Si at around 480 $cm^{-1}$. These room-temperature spectral positions are shown in Table 2, which is adapted from Ref. [23], which also has multiple citations describing each vibrational mode.

**Table 2. Vibrational resonance modes in fused silica around room temperature [23].**

| Wavenumber (cm-1) | Wavelength (μm) | Description of vibrational Mode |
| --- | --- | --- |

| ~1070–1120 | ~8.9–9.35 | Asymmetric stretching of Si-O-Si (bridging oxygen (BO)) |
|---|---|---|
| ~950 | ~10.5 | Stretching of Si-OH |
| ~800 | ~12.5 | Symmetric stretching or bending of Si-O-Si |
| ~480 | ~21 | Bending (rocking) of Si-O-Si |

It is initially tempting to assign one Gaussian oscillator to each of the vibrational resonance modes in Table 2. However, we were unable to fully fit the experimental data (initially just at room temperature) using a single Gaussian oscillator per mode in Table 2, for a total of four oscillators.

Increasing the number of Gaussian oscillators from four to seven significantly improved the fit for our experimental data of standard-grade fused silica at room temperature, likely because some of the line shapes are asymmetric and are better described with multiple Gaussian oscillators than a single one [Fig. 2]. While the 5-Gaussian model showed some improvement over the 4-Gaussian model by adding an oscillator at ~300 $cm^{-1}$ [Fig. S4], mismatches persisted. In contrast, the 6- and 7-Gaussian models closely match the experimental Ψ data. Notably, the 7-Gaussian model's oscillator positions align closely with the eight-oscillator model by R. Kitamura et al. [25]. We used three oscillators for the 1070–1120 $cm^{-1}$ mode, one oscillator for the ~800 $cm^{-1}$ mode, two oscillators for the ~480 cm-1 mode, and a broad, shallow oscillator fixed at ~390 $cm^{-1}$. We note that due to the low impurity levels and OH content in all three grades of fused silica [Table 1], the Si-OH resonance mode at ~950 $cm^{-1}$ was not observed in our experimental data, and therefore no Gaussian oscillator was assigned to this region.

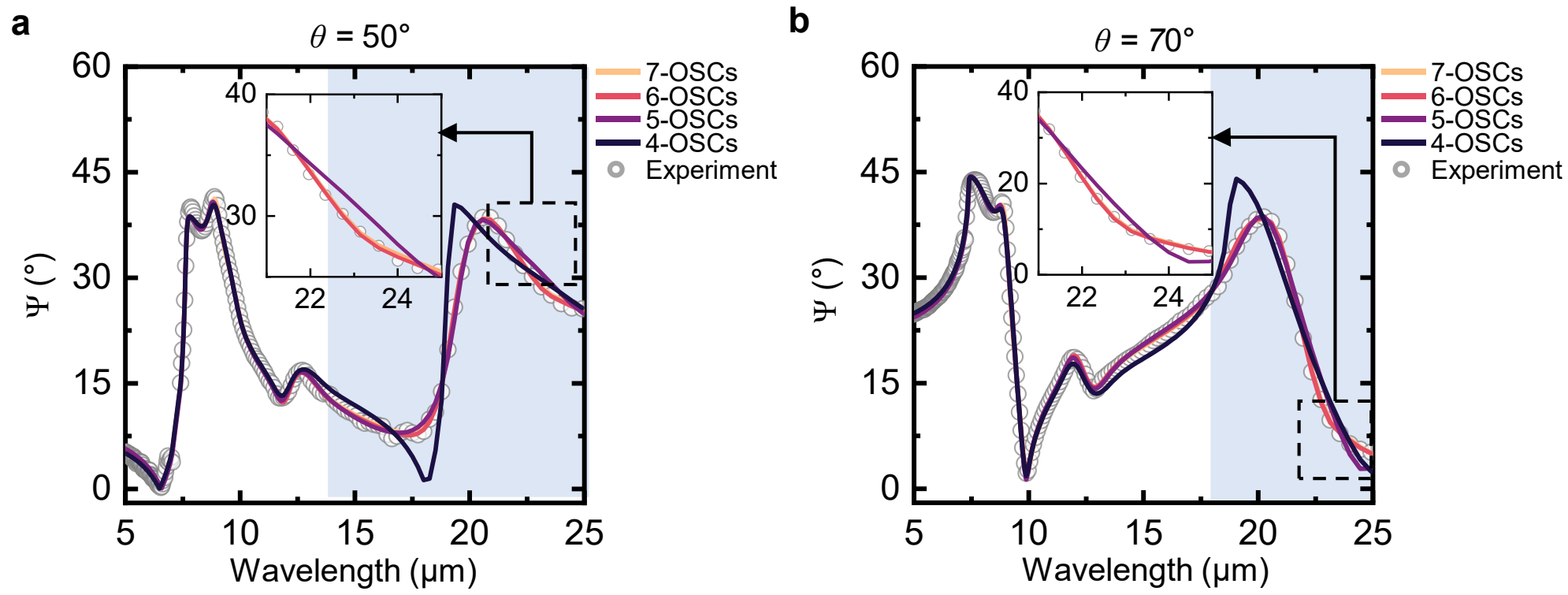

Figure 2. (a,b) Experimental Ψ data (symbols) for standard-grade fused silica at room temperature, with fits using 4, 5, 6, and 7 Gaussian oscillators (lines) at incident angles of (a) 50° and (b) 70°. In these plots, the 7-oscillator curve is mostly underneath the 6-oscillator curve. The blue shaded regions highlight wavelength ranges with the most significant differences between the 4-oscillator and 5-oscillator models and the other two models. Insets show zoomed-in plots of the dashed square regions, highlighting the deviation of the 5-oscillator model from the experimental data.

The 7-oscillator fit matches the experimental data well, but it is not clear whether each oscillator describes a separate physical resonance mode. In an extreme case, one can imagine fitting data with, say, 100 oscillators, which will result in extremely good fits for a given dataset, but many of the oscillators will not correspond to physical features of the sample—in some cases fitting to noise in the measurement. This is an example of overfitting. Thus, our goal is to create a model that uses as few oscillators as possible to prevent overfitting and maintain physical meaning, while still obtaining a high-quality fit to the measured

Ψ and Δ. We diagnose overfitting by measuring the sample over a range of temperatures, expecting that vibrational modes (spectral position, amplitude, and spectral width of each mode) should change predictably with temperature. The parameters of each oscillator were carefully optimized to balance fitting accuracy with physical relevance, minimizing the risk of overfitting while accurately capturing the temperature-dependent optical properties [5].

We found that our 7-oscillator model is an example of overfitting to the ellipsometry data, as shown in Fig. 3. The experimental Ψ and Δ data exhibit monotonic changes with increasing temperature, as expected [Figs. 3a and b]. But the fitted amplitude of two adjacent oscillators (oscillators #2 and #3 in Fig. 3c, bottom) has the opposite temperature-dependent trend, and the width of one of the oscillators (#2) does not change monotonically with temperature [Fig. 3e], both indicating a potential overfit. All other oscillator parameters versus temperature plots are shown in Fig. S6. We also performed another completely independent fit using 7 oscillators, which resulted in different oscillator parameters (though nearly identical complex refractive index), as shown in Fig. S7, indicating that there is not a unique solution using 7 oscillators.

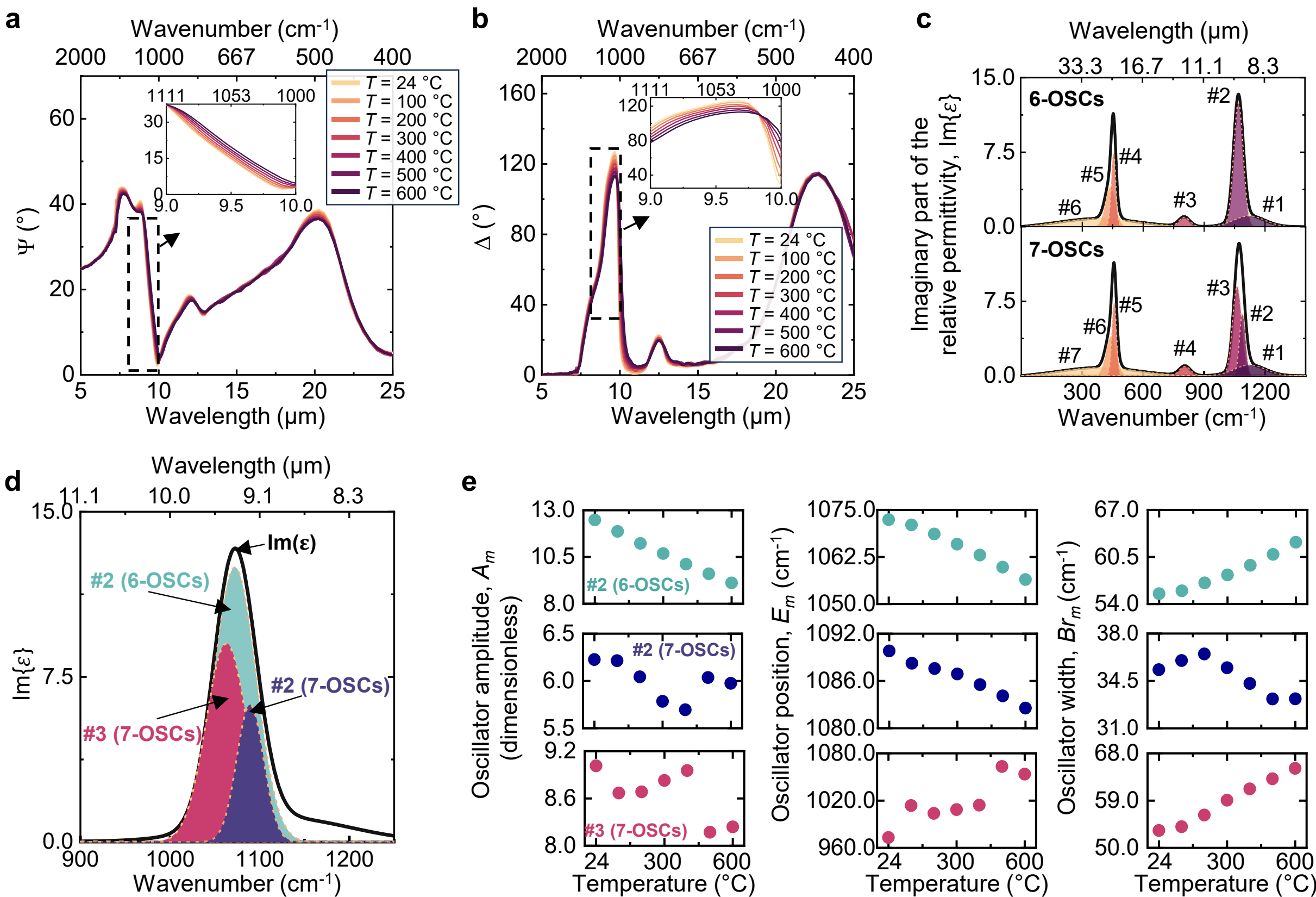

Figure 3. (a,b) Experimental (a) Ψ and (b) Δ for standard-grade fused silica as a function of temperature. Insets show zoom-in plots for the 9–10 μm region (dashed box), highlighting the monotonic temperature dependence of Ψ and Δ. (c, d) Imaginary part of the relative permittivity (Im{ε}) obtained using the (upper) 6- and (lower) 7-Gaussian oscillator models in (c). The difference is highlighted in (d), where the 7-Gaussian model splits the single oscillator (#2) from the 6-Gaussian model into two distinct oscillators (#2 and #3). (e) Temperature-dependent oscillator parameters—amplitude (left), spectral peak position (middle), and width (right) —for the highlighted oscillators in (d).

Hence, we reduced the total number of oscillators down to 6, shown in Fig. 4a, which still provides a very good fit to the experimental data [Fig. 2]. The temperature dependence of the parameters of these oscillators is shown in Figs. 4b–d, with the temperature dependence of $\varepsilon_\infty$ shown in Fig. S8. The results demonstrate that the 6-oscillator model exhibits consistent, monotonic temperature dependence for most oscillator parameters across all grades of fused silica samples [Figs. 4b-d].

In polar materials, phonon modes are expected to red-shift, broaden, and reduce in amplitude as temperature increases due to thermal expansion and anharmonic effects [26–28,43,44]. Anharmonicity, which arises from deviations from harmonic lattice vibrations—where atoms oscillate symmetrically around their equilibrium positions—becomes more pronounced at higher temperatures. This leads to enhanced phonon damping [43] and activates higher-order phonon-phonon interactions, such as three-phonon and four-phonon processes [26–28,43,44].

In our 6-oscillator fit in Fig. 4, oscillators #2, #3, and #5 have the expected temperature-dependent trend. Oscillator #6 is a low-amplitude, position-fixed, and broad oscillator that has more fitting noise. Oscillators #1 and #4 also show a monotonic temperature-dependent trend as expected, but with amplitudes that increase with temperature, which disagrees with the expected behavior based on thermal expansion and anharmonicity observed in oscillators #2, #3, and #5. This suggests that oscillators #1 and #4 on their own do not represent distinct vibrational modes. Instead, oscillators #1 and #2 together account for the asymmetric stretching of the Si-O-Si mode, and oscillators #4 to #6 together account for the bending (rocking) of the Si-O-Si mode (also see Supplemental Section V). In both of these cases, the line shape is asymmetric and cannot be fitted with any one symmetric oscillator function (e.g., Gaussian or Lorentzian). It is possible that more-complex oscillator models can be used to better represent the asymmetric line shapes of the physical resonance modes using only a single oscillator per mode but, for this paper, we decided to avoid the extra complexity of asymmetric oscillators.

The 6-oscillator model achieves a balance between accurate fit to the ellipsometry data and physical relevance, without introducing unnecessary complexity.

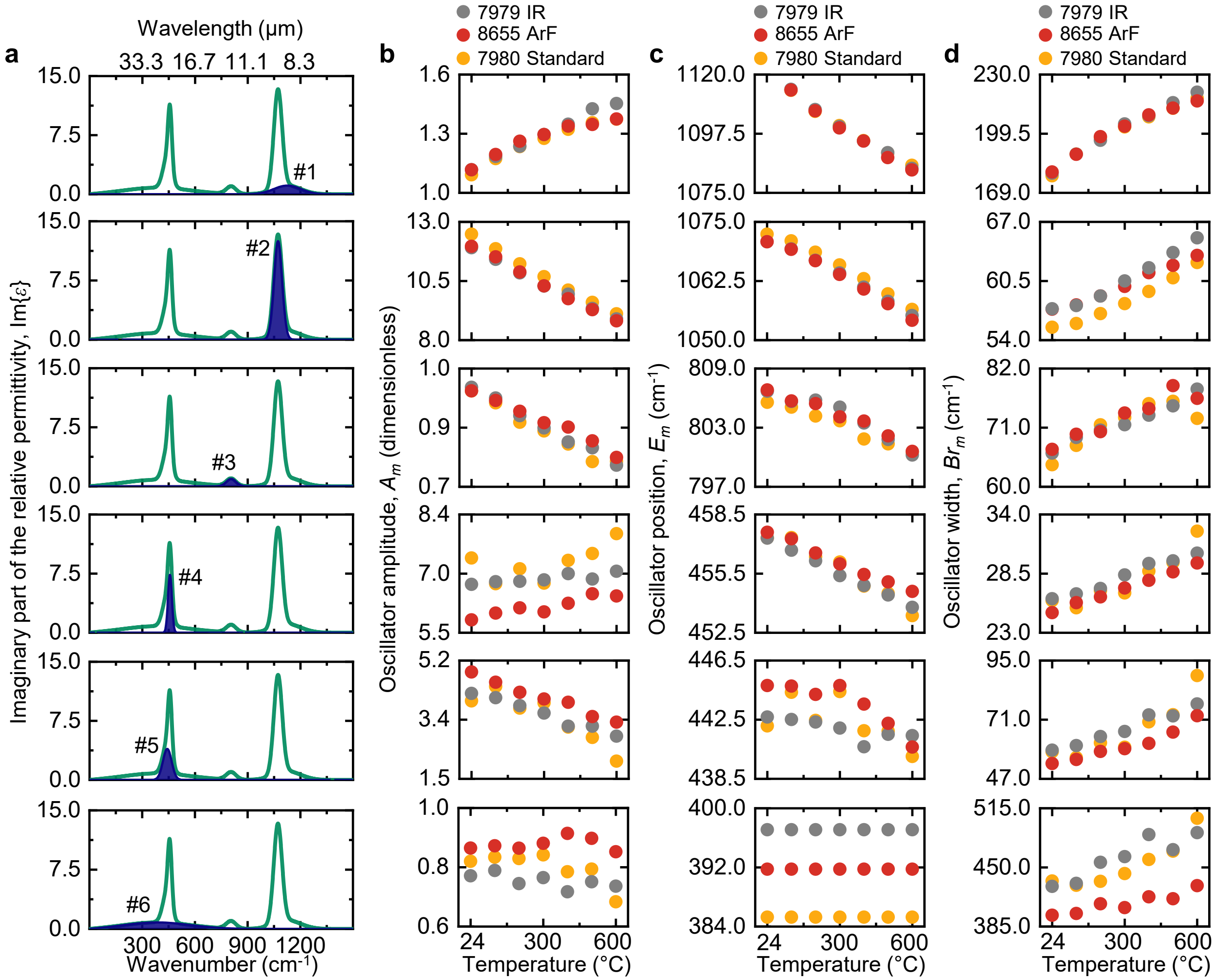

Figure 4. Temperature dependence of the 6-Gaussian oscillator model parameters used to fit the ellipsometric data. (a) Each row represents one of the six oscillators (#1–#6), with the contribution from a specific oscillator highlighted in blue. (b–d) Plots of oscillator parameters; (b) amplitude, (c) spectral peak position, and (d) spectral width as a function of temperature for each grade of fused silica. Note that the fitting for each temperature was performed independently and manually using the WVASE program across all temperature points.

Because the resulting dataset of temperature-dependent optical properties can be useful for design and simulations of mid-infrared optical components, especially at elevated temperatures, we worked to generate complete $n, \kappa$ datasets as a function of temperature, included as a supplementary data set. However, we found that to provide accurate $n$ and $\kappa$ data across the entire 5–25 μm range, we could not rely exclusively on ellipsometry data [Figs. 1–4], because the $\kappa$ value in the 5–6.5 μm range is too small to be reliably measured with ellipsometry, at least in our measurement configuration. To address this limitation of the ellipsometry measurement, we performed temperature-dependent transmission measurements through the full 500-μm-thick double-side-polished wafers, and used the measurements to add a low-amplitude Gaussian oscillator (< 1/10 the magnitude of the 6 oscillators in Fig. 4) with sufficiently broad width to each model. This broadness allows the oscillator to influence the highly transparent region, even though its position lies outside that region [Figs. S9–11]. We refer to this final model as the 6+1 Gaussian oscillator model. The resulting tabulated $n, \kappa$ data includes this correction, and is valid across the entire 5–25 μm range (Fig. 5 and the supplementary dataset).

The resulting complex refractive indices are very similar between the different grades, although our measurements are sufficiently sensitive to quantify the minor differences between them. When zooming into an individual resonance, the temperature dependence can be clearly observed. For example, the feature around 9 μm (insets in the upper panels of Fig. 3) due to Si-O-Si stretching [22] gradually decreases in amplitude and shifts to longer wavelengths as the temperature increases [26,27].

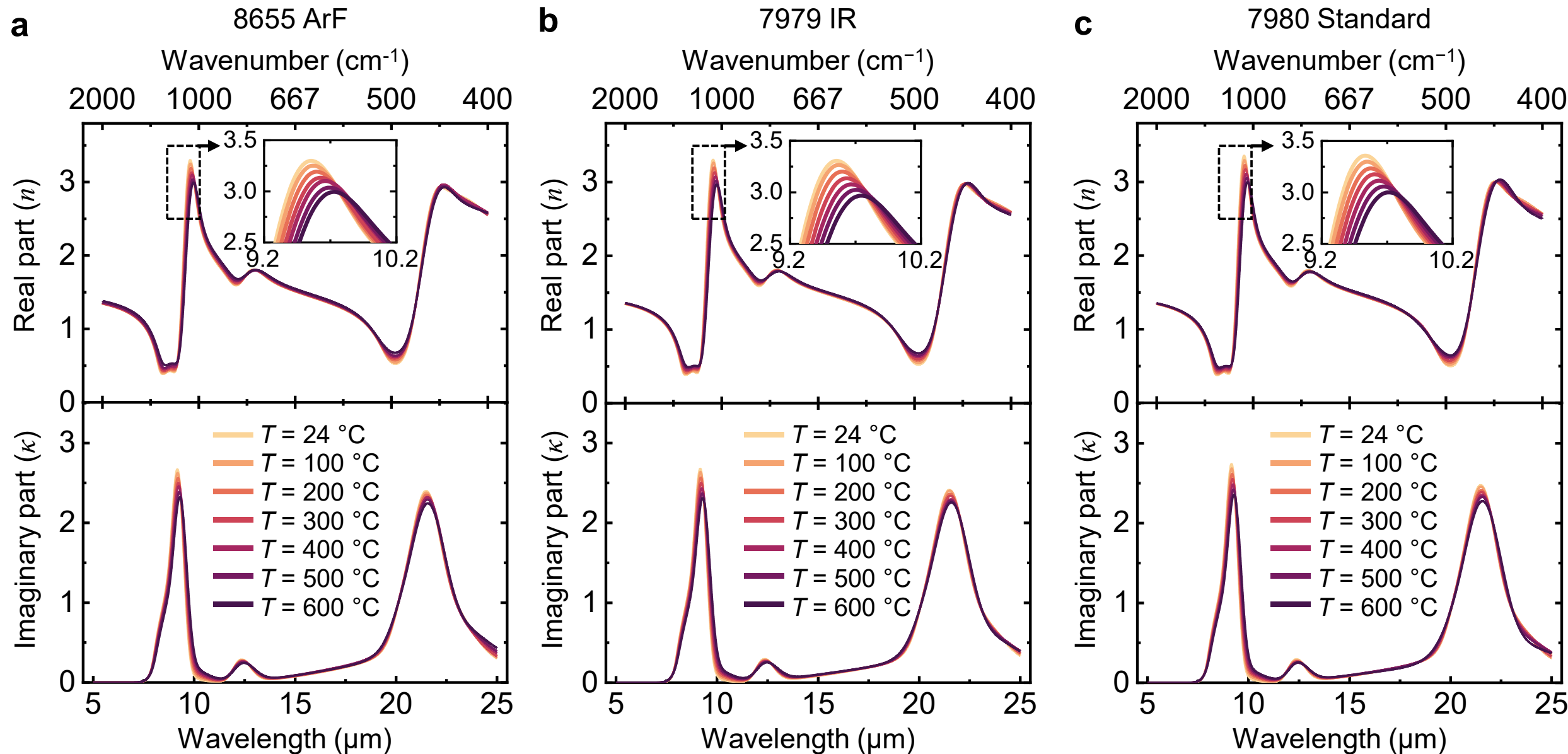

Figure 5. (a–c) Temperature-dependent complex refractive indices for different grades of fused silica: (a) 8655 ArF Grade (< 10 ppb metallic impurities, < 1 ppm OH); (b) 7979 IR Grade (< 100 ppb metallic impurities, similar OH as (a)); (c) 7980 Standard Grade (< 1000 ppb metallic impurities, 800–1000 ppm OH). Insets show the monotonic changes in the complex refractive index with increasing temperature.

As a comparison, we plotted the final room-temperature complex refractive indices for all grades of fused silica, together with the data from Popova et al. (who used an 8-oscillator model) [35] in Fig. S12.

## 4. Conclusion

We systematically characterized the optical properties of different grades of fused silica over a range of temperatures (24–600 °C) using Fourier-transform variable-angle spectroscopic ellipsometry (VASE). By incorporating the temperature dependence into our fitting process, we achieved balance between fitting accuracy and physical meaning of the oscillators, accurately capturing the temperature dependence of the experimental data. The result is a high-quality dataset of the temperature-dependent complex refractive indices of each grade of fused silica across the wavelength range of 5 to 25 μm, ready to be used for high-accuracy modeling of infrared absorbers, thermal emitters, and other photonic structures that operate at elevated temperatures. More broadly, the incorporation of the temperature degree of freedom for materials with well-behaved temperature dependence can improve the accuracy of many optical measurements, from FTIR absorption spectroscopy to VASE to Raman spectroscopy.

## Acknowledgements

This work was supported by the Office of Naval Research (N00014-20-1-2297), as well as the Center for Semiconductor Thermal Photonics supported by the UW-Madison Office of the Vice Chancellor for Research with funding from the Wisconsin Alumni Research Foundation. The authors also gratefully acknowledge use of facilities and instrumentation at the UW-Madison Wisconsin Centers for Nanoscale Technology (wcnt.wisc.edu) partially supported by the NSF through the University of Wisconsin Materials Research Science and Engineering Center (DMR-1720415).

## Data availability

Data presented in the main text and supplementary information are open access and can be found on Zenodo: https://zenodo.org/records/15178466

*Supplementary Information for:*

# Preventing overfitting in infrared ellipsometry using temperature dependence: fused silica as a case study

**Shenwei Yin[1, †], Jin-Woo Cho[1, †], Demeng Feng[1], Hongyan Mei[1], Tanuj Kumar[1], Chenghao Wan[1], Yeonghoon Jin[1], Minjeong Kim[1], and Mikhail A. Kats[1*]**

*[1]Department of Electrical and Computer Engineering, University of Wisconsin-Madison, Madison, Wisconsin 53706, USA.*
*[†]These authors contributed equally.*
**mkats@wisc.edu*

Supplement Section I-IX
Supplementary References

## Section I. Ellipsometer settings

For all of the measurements in this paper, we used the following settings on the J. A. Woollam IR-VASE instrument:

"Spectra/rev": 15, which is the number of azimuthal orientations of the compensator used during a measurement. The compensator introduces a wavelength-independent quarter-wave phase shift.

"Scans/spectrum": 30, which is the number of FTIR scans acquired at each compensator position.

Resolution: 16 $cm^{-1}$

"Bandwidth": 0.0005 μm, which is a mathematical smoothing parameter defined by Woollam.

## Section II. Pseudo-Brewster angle calculations

The pseudo-Brewster angle, $\phi_B$, is the angle at which p-polarized reflectance from a material with a complex refractive index is minimized. If the material is lossless, the pseudo-Brewster angle is the same as the Brewster angle.

Here, we calculate $\phi_B$ using the analytical solution in [S1], assuming light is incident from a material with refractive index $n_0$ (here, $n_0 = 1$) onto a material with complex refractive index $n + i\kappa$. $\phi_B$ can be found by solving the following equation:

$$\eta^3 - 3\alpha\eta + 2\beta = 0, \tag{S1}$$

where

$$\eta = \cot^2 \phi_B + \frac{1}{3}, \tag{S2}$$

$$\alpha = \left[\frac{n_0^2}{n^2 + \kappa^2}\right]^2 + \frac{1}{9}, \tag{S3}$$

$$\beta = \left[\frac{n_0^2}{n^2 + \kappa^2}\right]^3 \cdot \frac{n^2 - \kappa^2}{n^2 + \kappa^2} + \frac{1}{27} \tag{S4}$$

Plugging in the $n + i\kappa$ values for standard-grade fused silica at room temperature (from our fits in the main text), we calculated $\phi_B$, plotted in Fig. S1. Other than near resonances, $\phi_B$ falls into the 50° - 70° range across our wavelength range of interest, so we select these angles for our IR-VASE ellipsometry measurements to maximize the measured depolarization in the experiment.

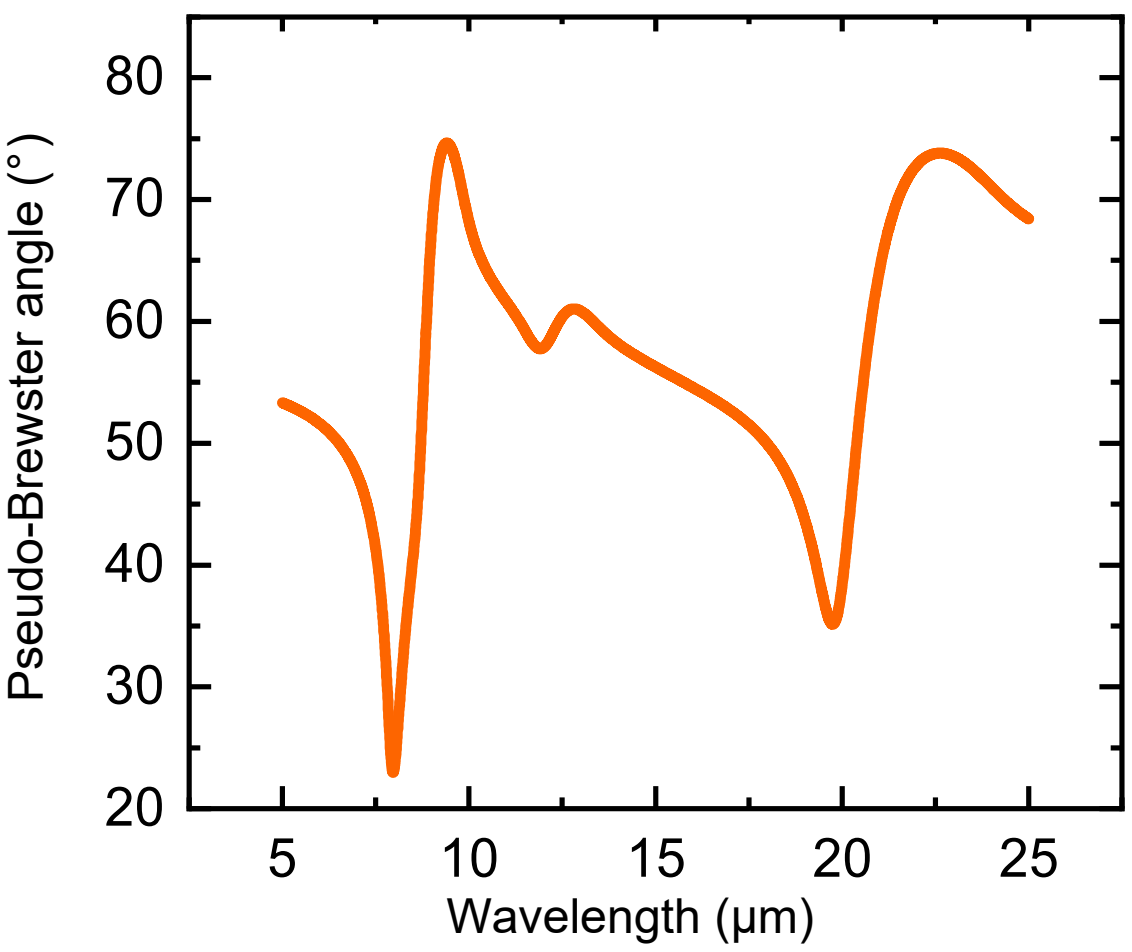

Figure S1. Calculated pseudo-Brewster angle for standard-grade fused silica at room temperature, using the refractive index obtained in this study.

**Section III. Error analysis for experimental data**

To demonstrate the uncertainty in our measurement, we performed 15 consecutive ellipsometry measurements on a standard-grade fused silica wafer, at an incident angle of 70° at room temperature (24 °C). In Fig. S2, we plotted the range of Ψ that we measured across these 15 measurements at each wavelength. The uncertainty is the largest in the wavelength range of 12.5–17.5 μm, and above 22.5 μm.

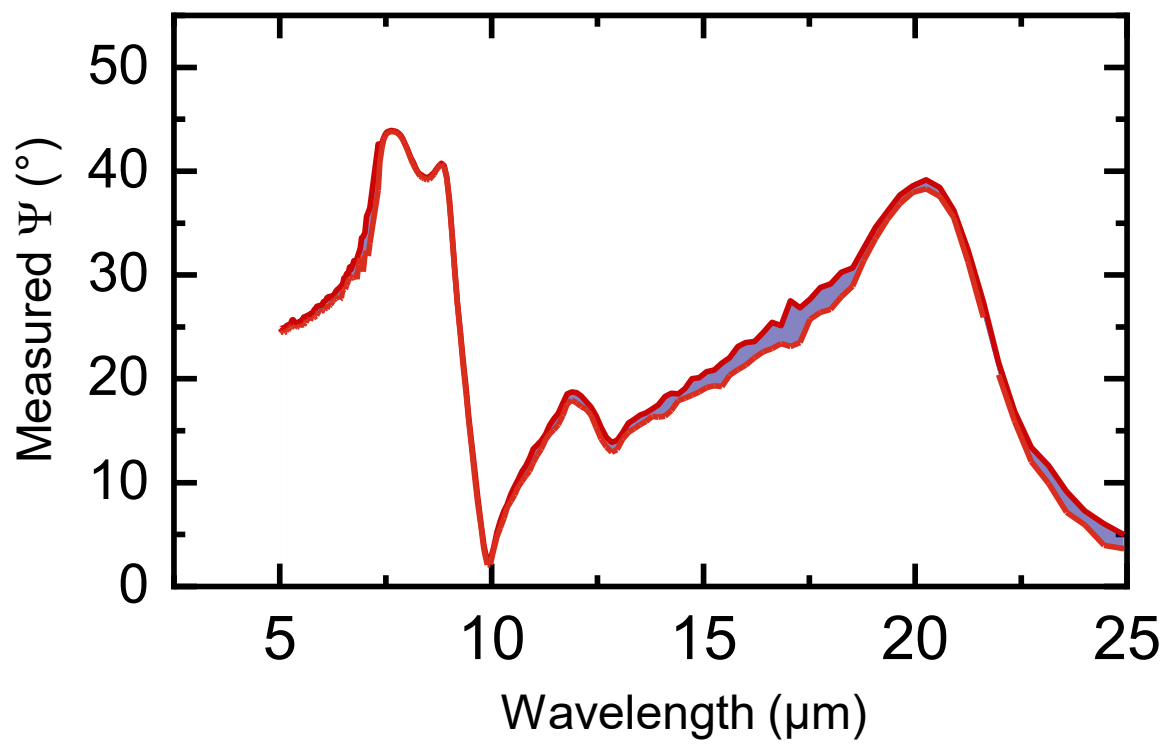

Figure S2. Experimental Ψ data from 15 consecutive ellipsometry scans at an incidence angle of 70° for standard grade fused silica (Corning 7980) at room temperature (24 °C). Blue shaded regions highlight the regions of short-term uncertainty in the measurement.

**Section IV. Comparison between Gaussian and Lorentzian oscillators**

In this section, we describe our attempt to fit the experimental data only using Lorentz oscillators (as opposed to Gaussian oscillators, as in the main text). A Lorentz oscillator is defined as Eqn. (S5) [S2]:

$$\varepsilon_{n_{Lorentz}} = \frac{A_n Br_n E_n E}{E_n^2 - E^2 - iBr_n E} \quad \text{(S5)},$$

where $\varepsilon_{n_{Lorentz}}$ is the complex relative permittivity, $A_n$ is the amplitude of the oscillator (dimensionless), $Br_n$ is the broadening of the oscillator (cm$^{-1}$), $E_n$ is the spectral position of the oscillator in spectroscopic wavenumber (cm$^{-1}$), and $E$ is the wavenumber of light (cm$^{-1}$). In Fig. S3, we compare the fitted results using only Lorentzian and Gaussian oscillators (six oscillators each, all of the same type), as well as a mixed case in which the three short-wavelength Gaussian oscillators from Fig. 4 in the main text were replaced with Lorentzian oscillators. The results show that the model composed entirely of Gaussian oscillators provides a better fit to the experimental data.

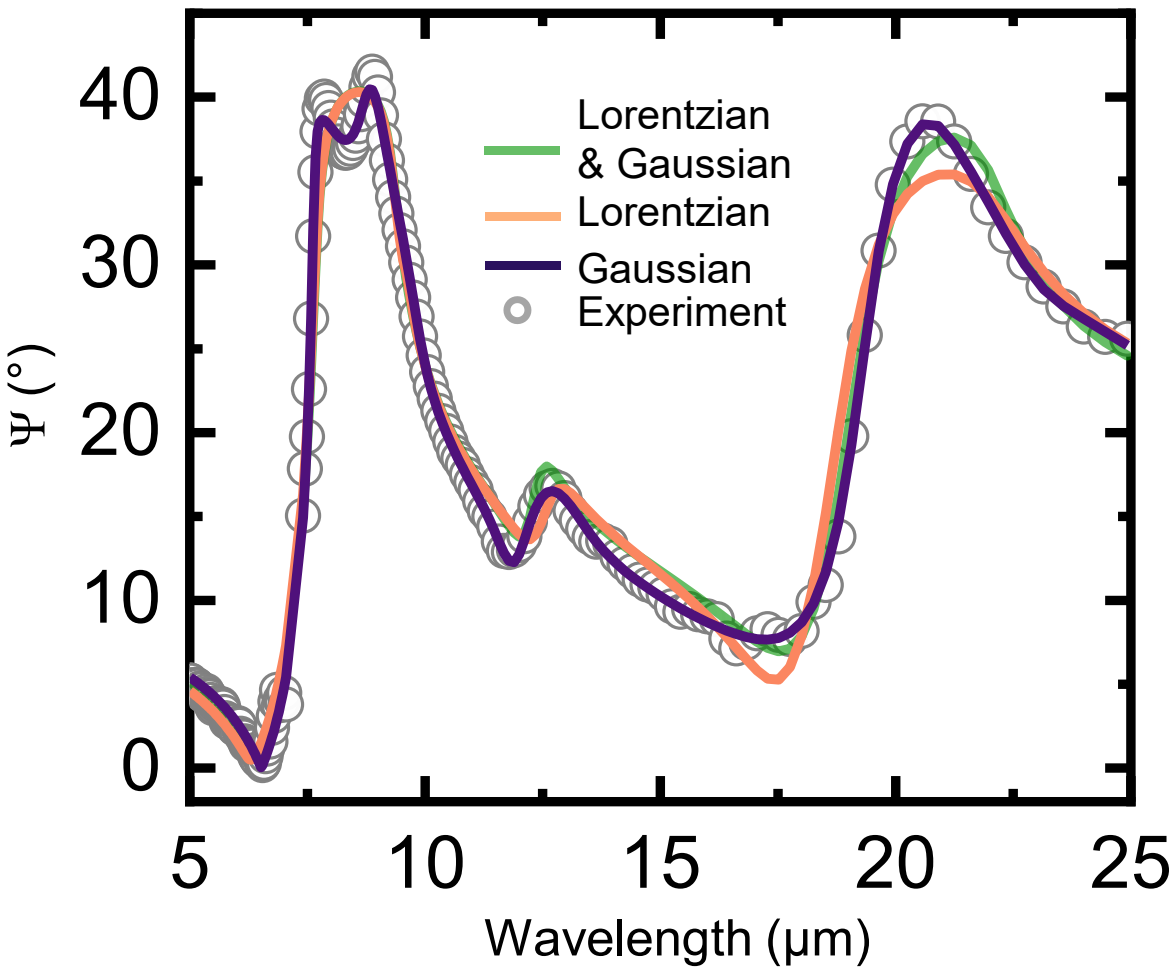

Figure S3. Experimental Ψ data (symbols) for standard-grade fused silica at room temperature, with fits (lines) using Lorentzian, Gaussian, and a combination of the two oscillator models. We replaced three short-wavelength Gaussian oscillators from Fig. 4 in the main text with Lorentzian oscillators. The fitting quality deteriorates in regions where Lorentz oscillators are used. And the fitting quality even gets slightly worse in the longer wavelength range, where Gaussian oscillators used in the main text remain unchanged. Therefore, the Gaussian model provides a better fit to the experimental data.

## Section V. 5-Gaussian oscillator model

Here we present the model comprising 5 Gaussian oscillators, as presented in Fig. 2 in the main text. Figure S4 shows the imaginary part of the relative permittivity (Im{ε}) obtained using 5-Gaussian oscillator model.

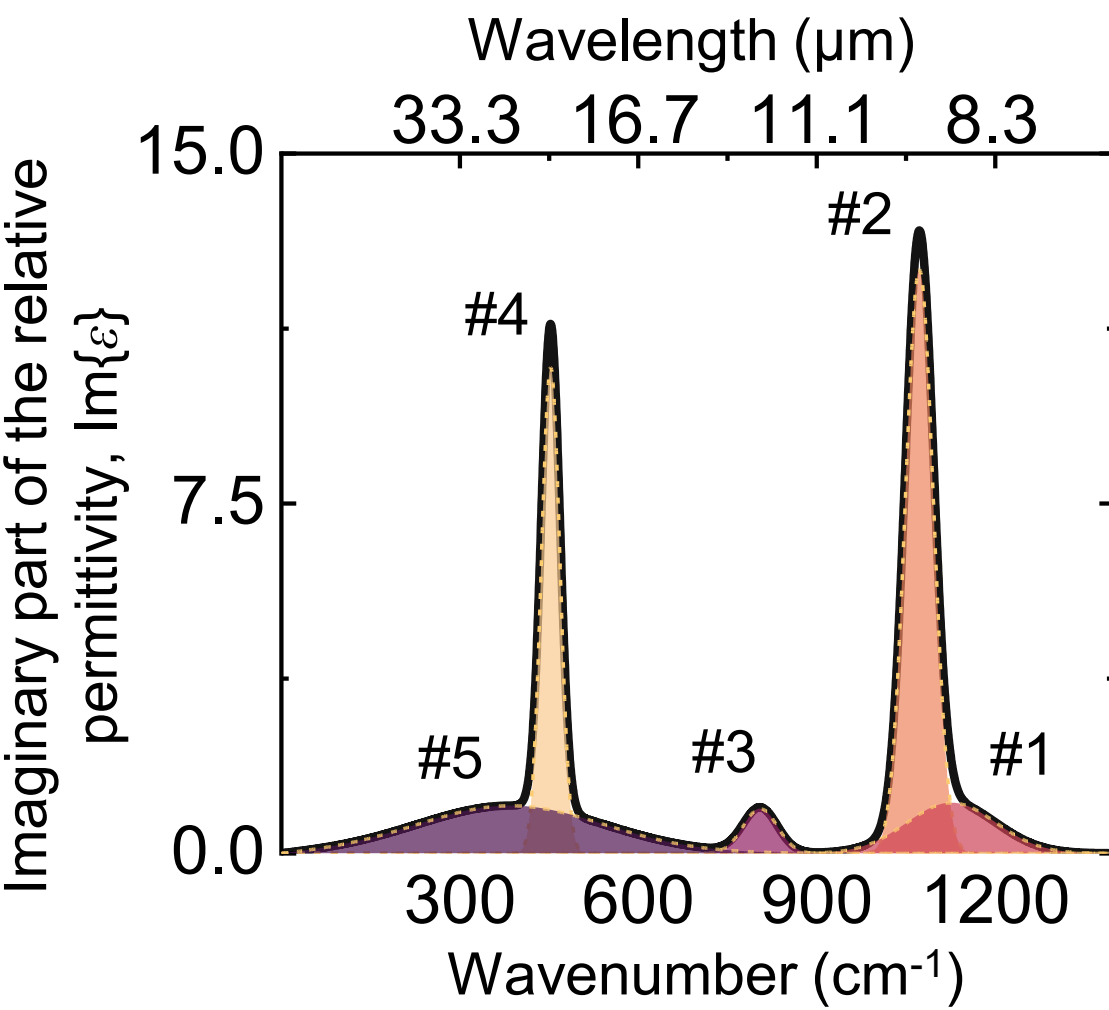

Figure S4. Imaginary part of the relative permittivity (Im{$\varepsilon$}) obtained using 5-Gaussian oscillator models.

Figure S5 shows the fitting parameters used in this 5-Gaussian oscillator model that shown in Fig. S4 as a function of temperature for all three grades. We collapse oscillators #4 and #5 in Fig. 4a into a single oscillator, keeping all other oscillator parameters identical. We observe, for example, that the differences between the oscillator amplitudes (Am) for oscillator #4 in Fig. 4(b) are quite large, but that spread is reduced for the combined oscillator in Fig. S5.

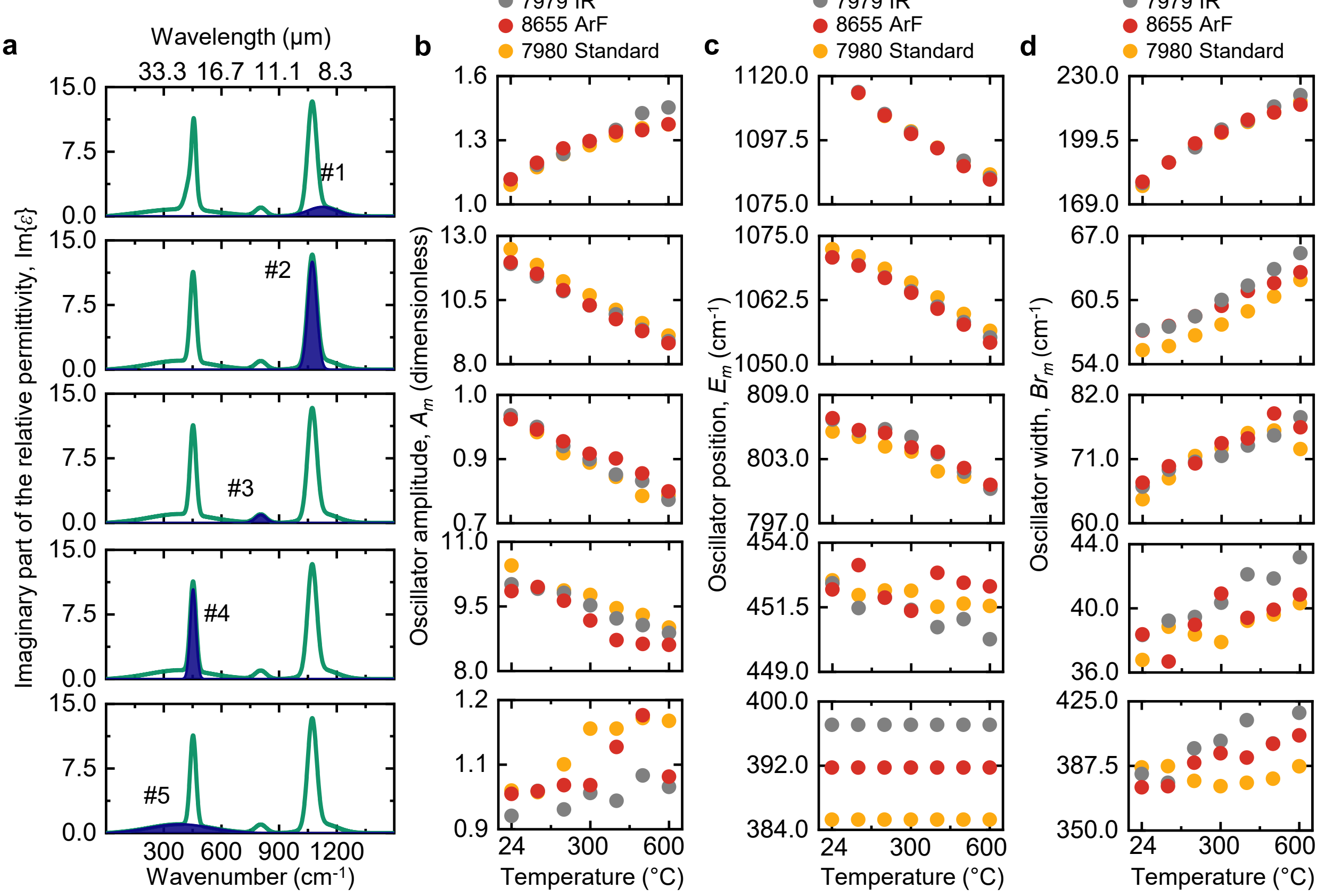

Figure S5. Temperature dependence of the 5-Gaussian oscillator model parameters used to fit the ellipsometric data. (a) Each row represents one of the five oscillators, with the contribution from a specific oscillator highlighted in blue shaded area. (b–d) Plots of oscillator parameters; (b) amplitude, (c) spectral peak position, and (d) spectral width as a function of temperature for each grade of fused silica. Note that the fitting for each temperature was performed independently and manually using the WVASE program across all temperature points.

## Section VI. Fitting parameters of 7-Gaussian oscillator model

We performed two independent fits on standard-grade fused silica using seven Gaussian oscillators. As discussed in the main text and shown in Fig. S5, the amplitudes of the two adjacent oscillators (#2 and #3) have exactly opposite temperature-dependent trends, indicating a potential overfit. A second attempt using seven Gaussian oscillators to fit the data is shown in Fig. S6; the oscillator parameters do not exhibit a clear, monotonic temperature-dependent trend. Both attempts demonstrate that the 7-Gaussian oscillator model will overfit the experimental data.

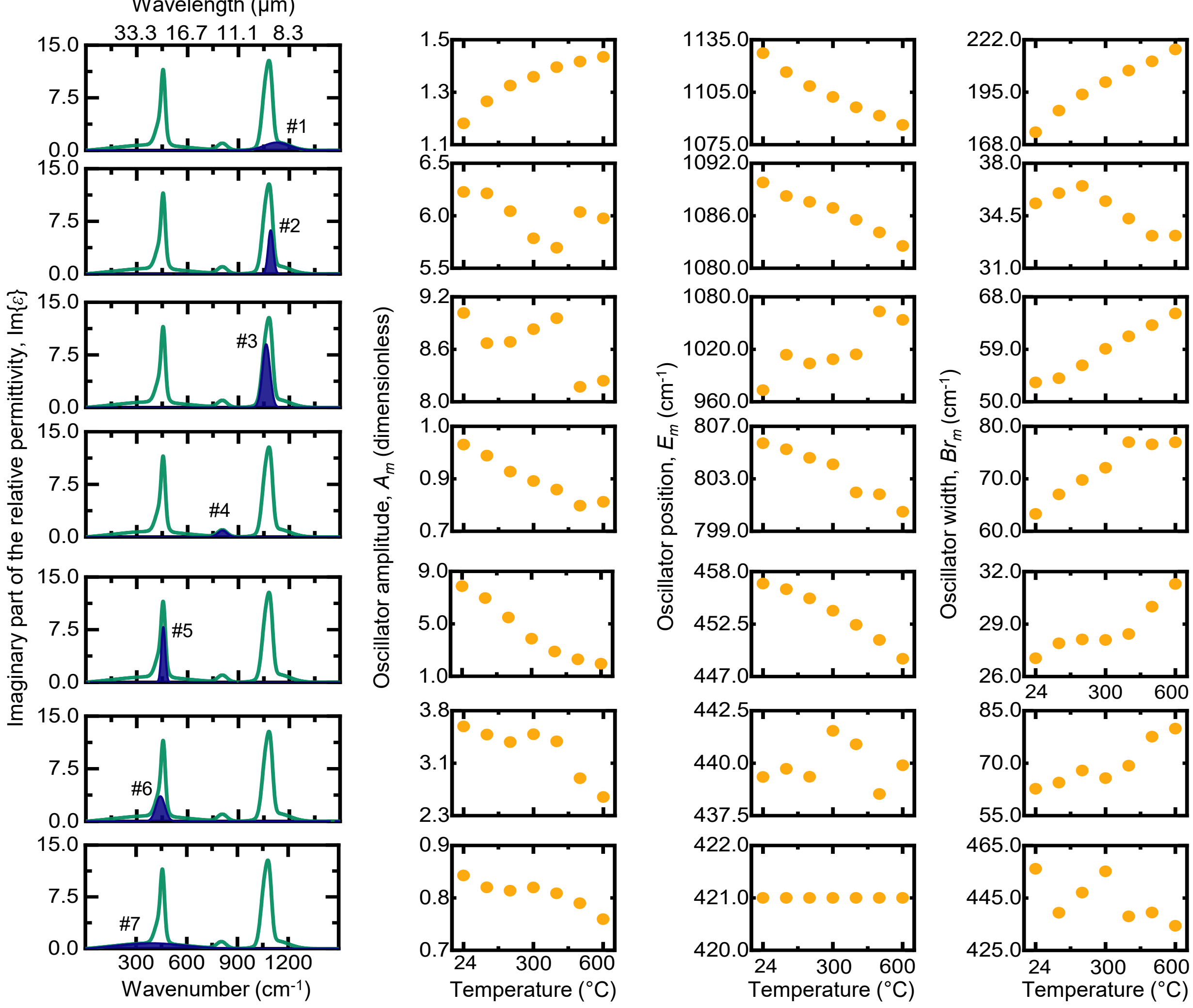

Figure S6. Temperature dependence of the 7-Gaussian oscillator model parameters used to fit the elipsometric data. This is the model presented in Fig. 3 in the main text. (a) Each row represents one of the six oscillators, with the contribution from a specific oscillator highlighted in blue shaded area. (b–d) Plots of oscillator parameters; (b) amplitude, (c) spectral peak position, and (d) spectral width as a function of temperature for each grade of fused silica. Note that the fitting for each temperature was performed independently and manually using the WVASE program across all temperature points.

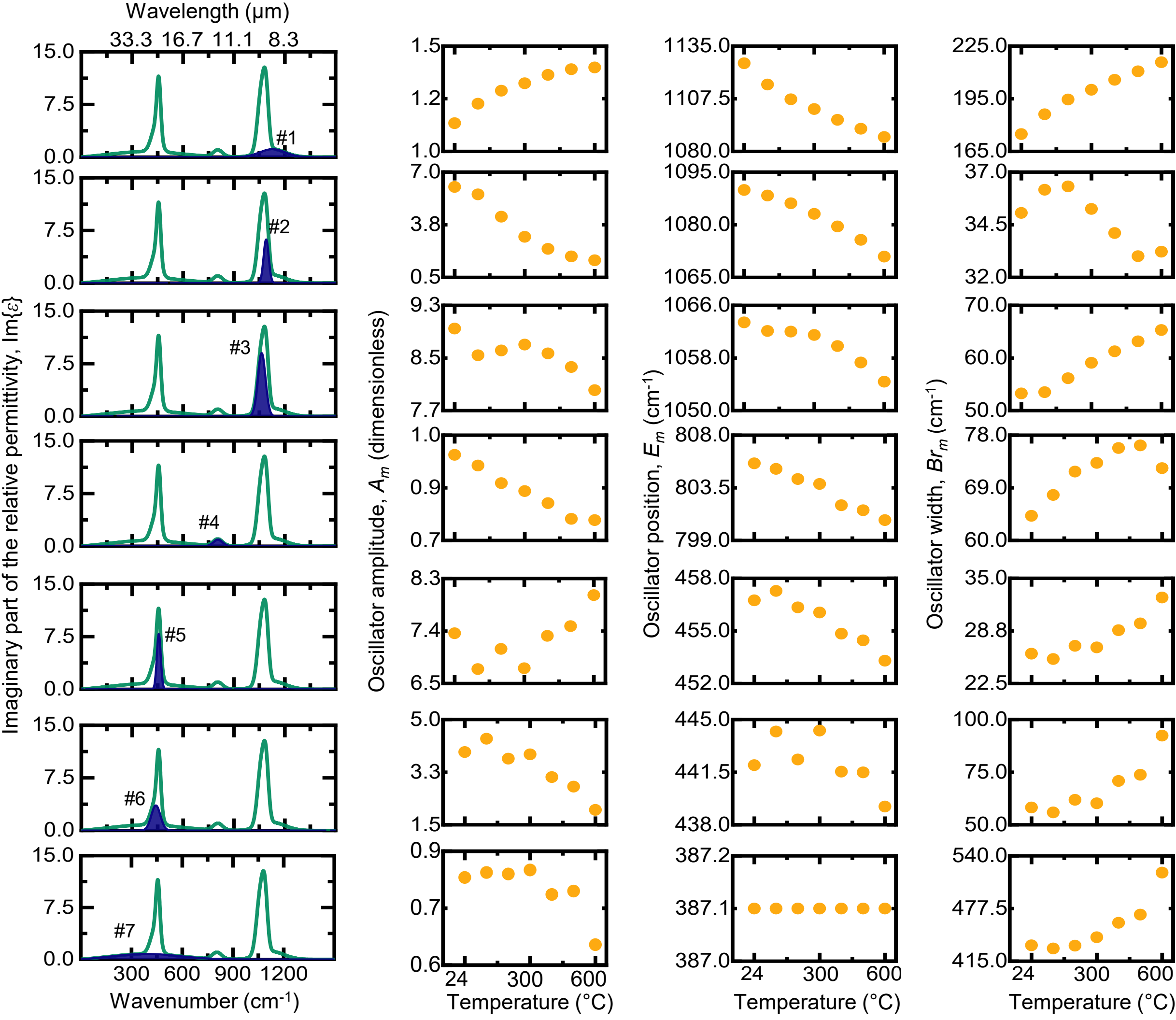

Figure S7. Second fit: Temperature dependence of the 7-Gaussian oscillator model parameters used to fit the elipsometric data. *Note that this fit is completely independent from the results in Figure S3*. While no contracting temperature-dependent trends were observed between the oscillators, the fitting parameters do not show consistent monotonic behavior with temperature, indicating possible overfitting. (a) Each row represents one of the six oscillators, with the contribution from a specific oscillator highlighted in blue shaded area. (b–d) Plots of oscillator parameters; (b) amplitude, (c) spectral peak position, and (d) spectral width as a function of temperature for each grade of fused silica. Note that the fitting for each temperature was performed independently and manually using the WVASE program across all temperature points.

## Section VII. Fitting parameter $\varepsilon_\infty$ versus temperature

We plot the fitting parameter $\varepsilon_\infty$ used in the 6-Gaussian oscillator model (the one presented in the main text, in Figs. 2-5) versus temperature for all three grades (Fig. S7). $\varepsilon_\infty$ also changes monotonically with temperature.

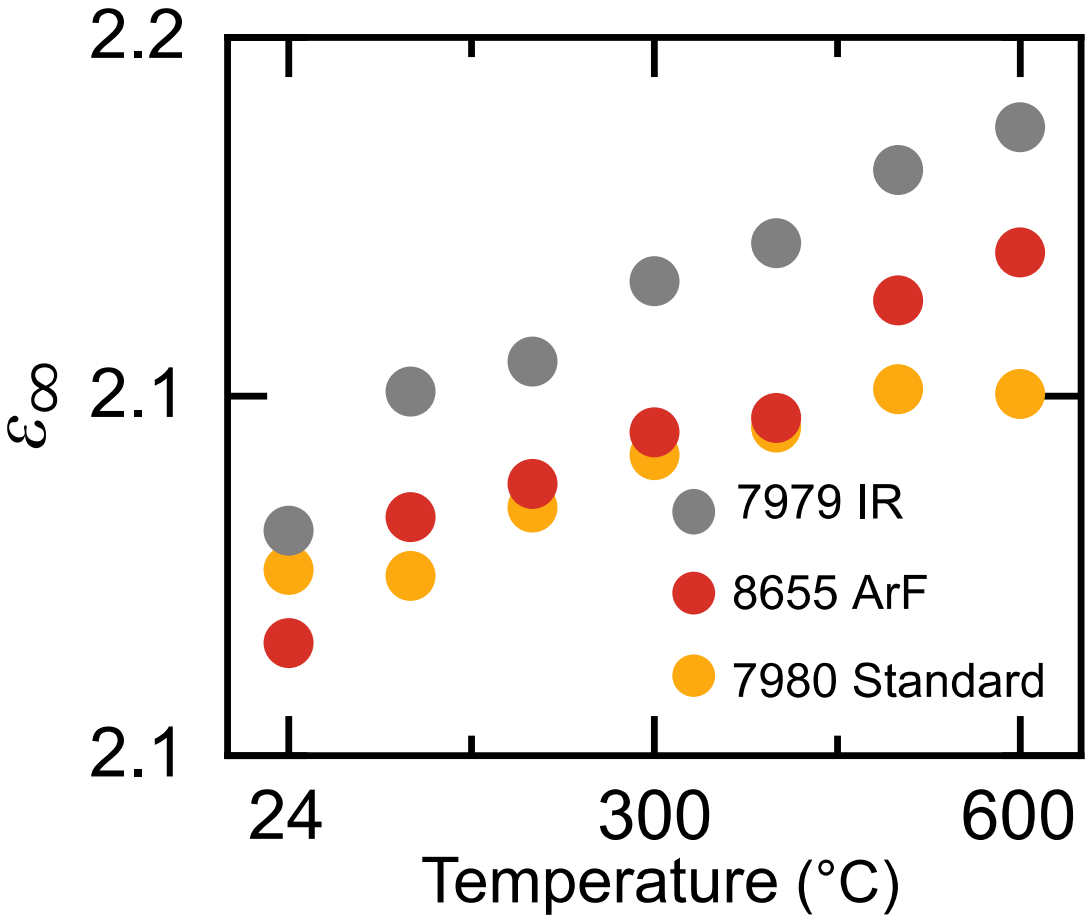

Figure S8. Temperature dependence of $\varepsilon_\infty$ of each grade of fused silica, using the 6-Gaussian oscillator model. The fitting parameter $\varepsilon_\infty$ also changes monotonically with temperature.

## Section VIII. FITR transmission measurements and refractive index calibration

To make sure that our tabulated $n, \kappa$ data is accurate in the short-wavelength range where fused silica has lower losses, we performed temperature-dependent Fourier transform infrared (FTIR) transmission measurements (Vertex 70, Bruker) using a liquid nitrogen-cooled photovoltaic mercury cadmium telluride (MCT) detector (D317-B, Bruker), as shown in Fig. S8.

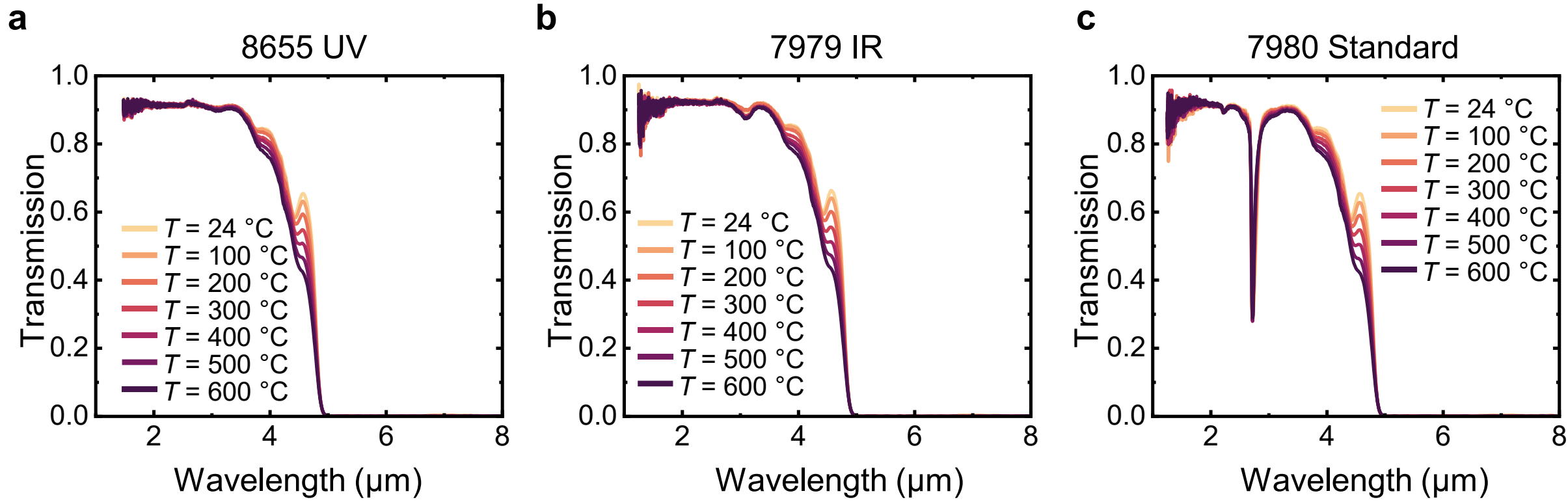

Figure S9. (a-c) Temperature-dependent FTIR transmission data for double-side polished wafers of three different grades of fused silica: (a) 8655 ArF Grade; (b) 7979 IR Grade; (c) 7980 Standard Grade. The differences in transmission values among the three grades are due to different impurity and OH content.

We found that the 6-oscillator model from the main text could not be used to reproduce the measured transmittance spectra from 3 to 6 µm in Fig. S8 (Fig. S9a and Fig. S9b), because ellipsometry of a semi-infinite sample (i.e., one interface) is not sufficiently sensitive to provide the necessary precision in the value of $\kappa$ for the material in this low-loss region.

We wanted to make sure that our reported datasets from 5 to 25 µm were highly accurate, including in the low-loss regions at shorter wavelengths, so we used the data in Fig. S8 to fit a seventh Gaussian oscillator with a very small amplitude to account for the actual loss in the material. This adjustment allowed us to achieve a good fit for both the transmission and ellipsometry data and get a more accurate refractive index dataset (Figs. S9c and S9d).

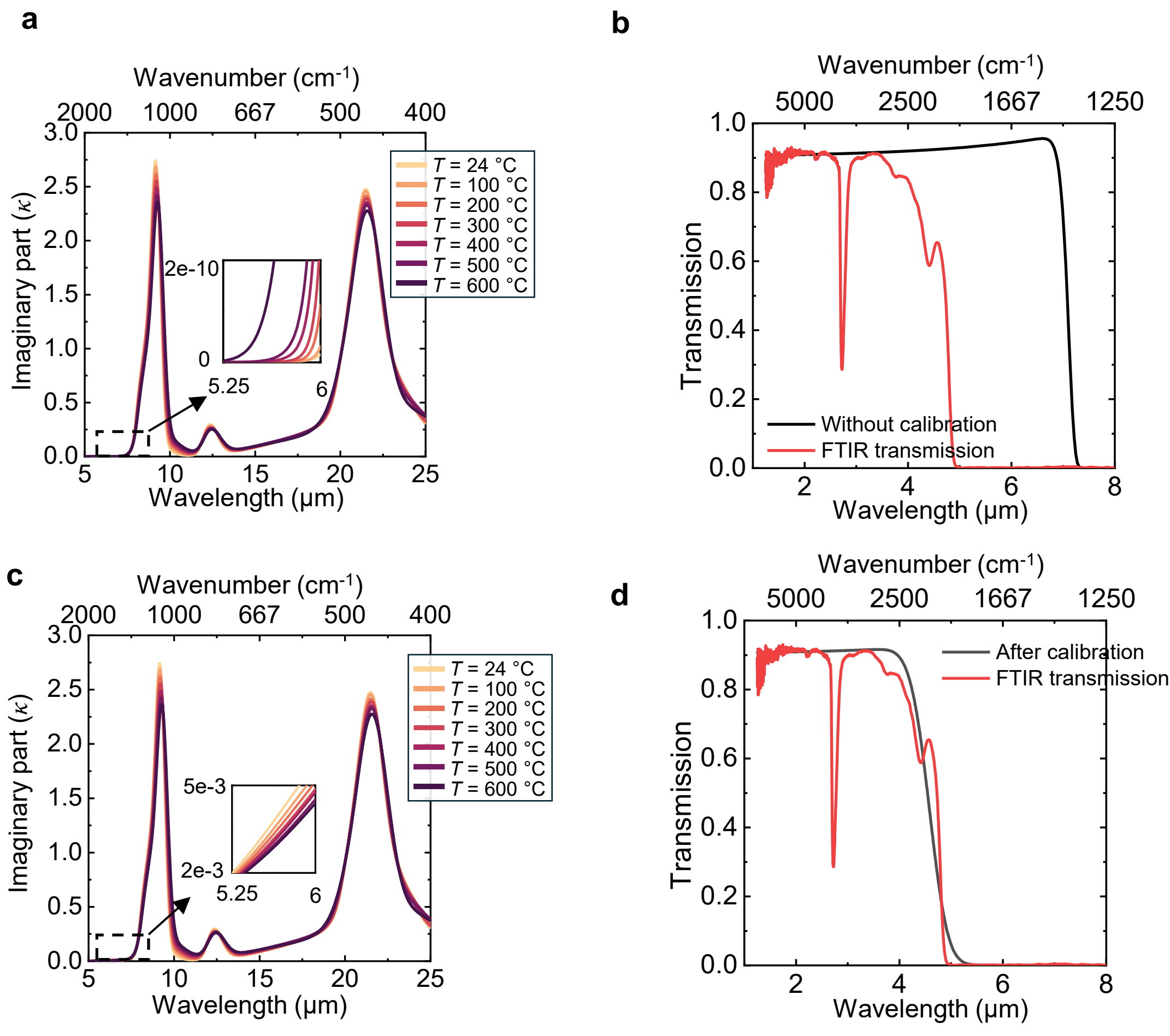

Figure S10. Comparison of the imaginary part of the complex refractive index of Corning 7980 Standard Grade fused silica before and after calibration using FTIR transmittance data. (a, b) Results before calibration, only using our fit to the ellipsometry data: (a) the temperature-dependent imaginary part of the complex refractive index. The inset shows a zoomed-in view highlighting the magnitude of the imaginary part. (b) Transmittance spectra at room temperature, calculated using the $n$ and $\kappa$ values from the six-Gaussian-oscillator model described in the main text. (c, d) Plots similar to (a) and (b), but after adding an additional oscillator, now matching the measured FTIR transmittance spectrum.

The resulting seventh oscillator parameters can be found in Fig. S10. Note that this added oscillator is only used to correct $\kappa$ in the spectral region where $\kappa$ is low. Therefore, the amplitude of this added oscillator is much lower than the values shown on Fig. 4, and the fitting error is high.

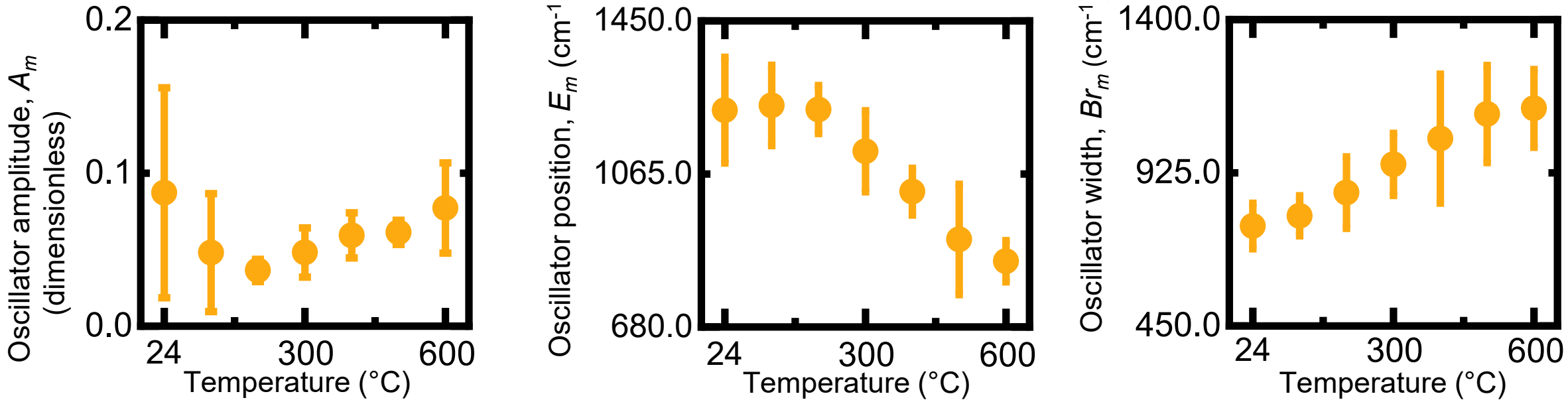

Figure S11. An additional small Gaussian oscillator in the short-wavelength range was introduced because, without it, the complex-refractive index data from ellipsometry could not reproduce the transmittance spectra in Figs. S8 and S9 (b and d). Here, we used 7980 standard grade fused silica as an example: we independently fitted the additional Gaussian oscillator three times at each temperature and plotted each parameter with error bars.

## Section IX. Comparison with literature data

In Fig. S11, we compared the complex refractive index of three different grades of Corning fused silica obtained from our study at room temperature with literature data from Popova et al [S3].

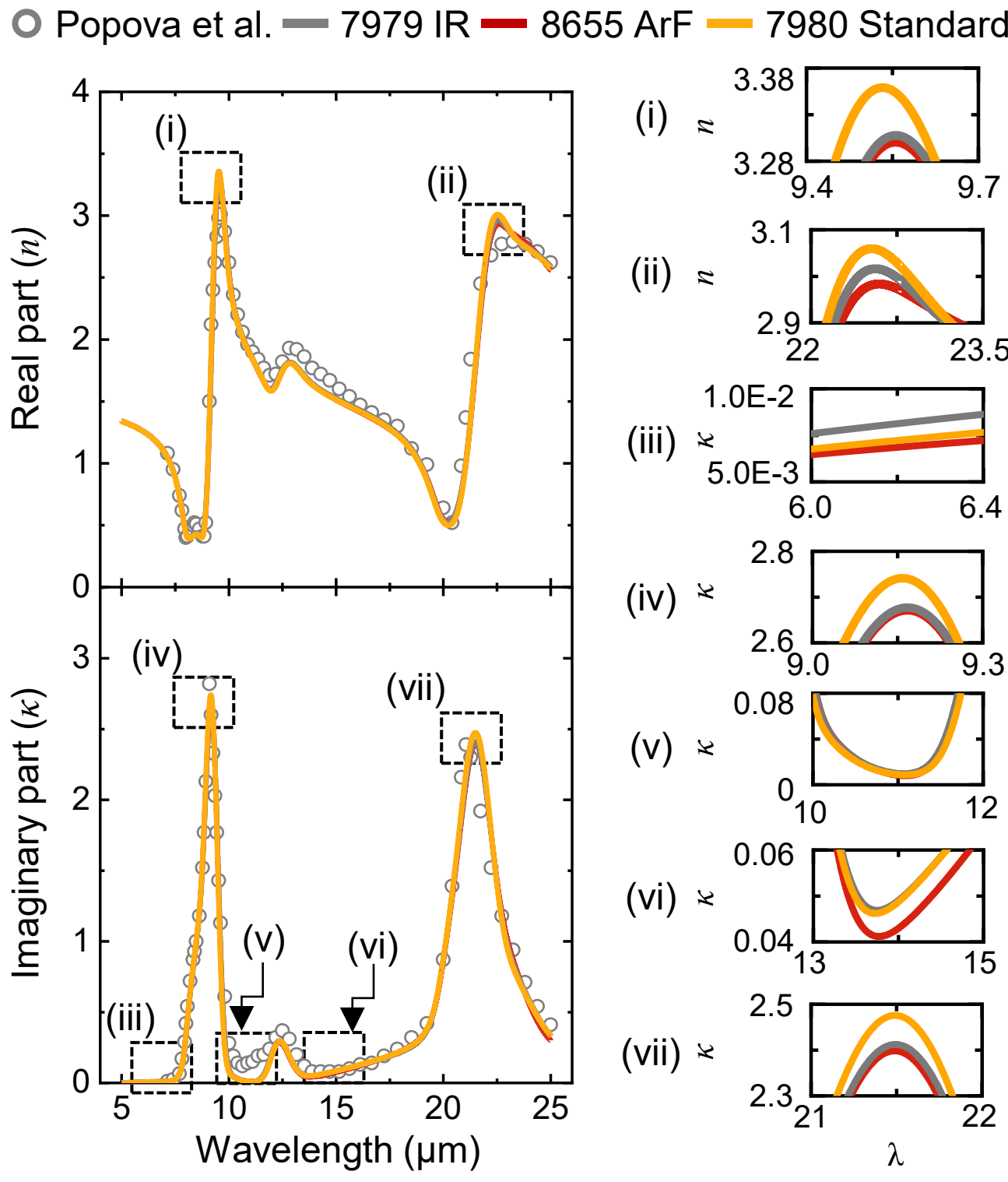

Figure S12. Comparison of the complex refractive index between our study and the data reported by Popova et al [3]. The zoomed-in areas in panel (i-vii) exclude the literature data,

providing a better comparison between the three different grades of fused silica utilized in our study. Notably, our measurements effectively captured the differences between the three different grades in plot (i), (ii), (iv), (vi), and (vii).

## Supplementary References